\documentclass[12pt]{iopart}
\usepackage{hyperref}
\usepackage{graphicx}
\usepackage{iopams}
\usepackage{cite}
	
\begin{document}
\title{Topological quantum pump of strongly interacting fermions in coupled chains}
\author{B A van Voorden and K Schoutens}
\address{Institute of Physics, University of Amsterdam,
		Science Park 904, Amsterdam, the Netherlands}
\ead{B.A.vanVoorden@uva.nl}

\begin{abstract}
	We analyze a tight binding model of two coupled chains with strongly interacting fermions. Depending on the parameter $w$, the many body lowest energy band consists of either single particles or bound pairs. A topological quantum pump can be created by periodically varying the coupling strengths under adiabatic conditions. We numerically show that the single particles and bound pairs are pumped in opposite directions along topologically non-trivial paths. The exact quantization of the pumped charge can be expressed by a (many body) Chern number. 
\end{abstract}

\maketitle

\section{Introduction}
In quasi-one-dimensional lattices with filled singe particle bands it is possible for charge transport to be quantized when the Hamiltonian is time-periodically varied under adiabatic conditions. This quantized charge transport can be related to a topological quantity, either the Chern number or the Zak phase, by the Berry phase of the time evolved wave function\cite{Berry1984,Xiao2010}. These systems are therefore named ``topological quantum pumps'' or ``Thouless pumps''\cite{Thouless1983}. Because this is a topological phenomenon, smooth variations of the pump path in parameter space have no influence on the total transported charge as long as the band gap remains open. This behaviour was first shown in the Su-Schrieffer-Heeger model of polyacetylene\cite{Su1979} and a generalization of this model by Rice and Mele\cite{Rice1982}. In recent years topological quantum pumps of non-interacting particles have been realized in experiments with ultracold atoms \cite{Lohse2016,Nakajima2016,Lu2016,Cooper2018}. With these versatile experimental setups it is in principle possible to tune the interactions between the particles, for example with Feshbach resonances, which opens up the possibility to experimentally investigate topological quantum pumps with interacting particles. Because of this there is a renewed theoretical interest on the topic of topological quantum pumps, specifically focusing on the role of interactions \cite{Rossini2013,Berg2011,Ke2017,Marques2017,Marques2017a,DiLiberto2016}.

Weakly interacting systems are well understood. As long as the ground state of the weakly interacting many body Hamiltonian is non-degenerate and separated with a finite energy gap from the excited states there will be quantized particle transport that can be expressed as a many body Chern number\cite{Niu1984, Niu1985,Xiao2010}. In contrast, there is still much that is not understood about the effect of strong interactions in topological quantum pumps. Recently, it was shown that in systems with cotranslational symmetry (invariance under the collective translation of the system) a many body Chern number could be defined that is linked to the displacement of many body Wannier states\cite{Qin2018}. This topological invariant can for example be used to characterize the topological features of bound states. In this article we investigate a topological quantum pump of coupled chains that have cotranslational symmetry and we will show that the strong interactions in our Hamiltonian lead to novel pump behaviour. Moreover, we will show that this invariant can not only be used to express the displacement of localized two body Wannier states, but also the current in a non-degenerate many body ground state.

\section{The model}
\begin{figure}[h]
	\centering
	\includegraphics[width=0.4\linewidth]{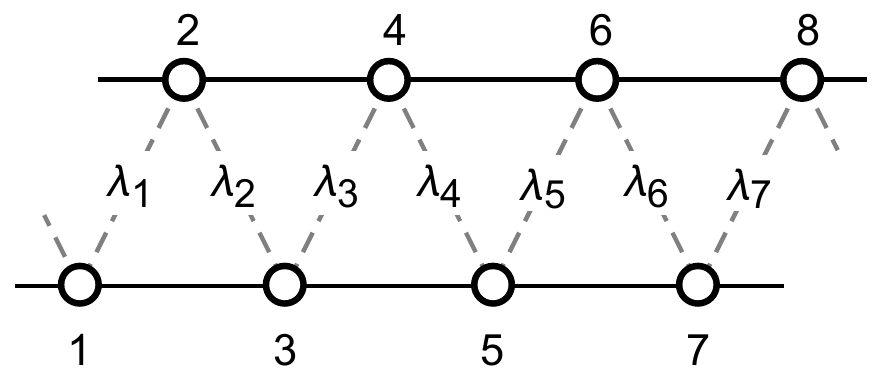}
	\caption{The 1D chain interpreted as two separate coupled chains in a zigzag pattern.}
	\label{fig:lambda1}
\end{figure}

Consider a tight binding model on a one dimensional chain of $L$ sites, divided into $p$ unit cells with $q$ sites each. We restrict ourselves to the cases where $L$ is even. On this chain $N$ spinless fermions are only allowed to hop to the next-nearest neighbour sites. Effectively, this model can be been seen as two separate chains lying next to each other in a zigzag pattern. One chain consists of all odd sites and the other of all even sites, see figure \ref{fig:lambda1}. Particles are only allowed to move along their own chain, so the number of particles on each chain, called $N_1$ and $N_2$, are separately conserved. The two chains are coupled by a local attractive interaction between particles in the different chains. Furthermore, particles hopping over an occupied site in the other chain pick up an extra factor of $\pm i$ compared to a hop over a non-occupied site. This extra phase can be interpreted as originating from flux tubes attached to the particles. Therefore the fermions effectively act as semions, which are anyons that get a factor of $i$ when exchanging two of them\cite{Wilczek1982}.

The Hamiltonian of the system is $H(\vec{\lambda}_L)=K(\vec{\lambda}_L)+V(\vec{\lambda}_L)$, with $K$ the kinetic hopping term and $V$ the density dependent potential. To characterize the strengths of these terms, we use the parameters $\vec{\lambda}_L=(\lambda_1,\lambda_2,\ldots,\lambda_L)$ where $\lambda_j$ gives the strength between sites $j$ and $j+1$. The origin of these $\lambda_j$'s is discussed below. Explicitly, the Hamiltonian is 

\begin{equation}
\eqalign{K = \sum_{j}^{}\lambda_{j+1}\,\lambda_{j}\, c_{j+2}^{\dagger} \,(p_{j+1}-in_{j+1})\,c_{j} + \rm{h.c.}\\
V = \sum_{j}^{}(\lambda_{j-1}^2 \,p_{j-1} + \lambda_{j}^2 \,p_{j+1})\,n_j.} \label{hamiltonian}  
\end{equation}

Here $c^{\dagger}_j$ and $c_j$ are the usual fermionic creation and annihilation operators, $n_j = c^{\dagger}_jc_j$ is the number operator and $p_j = (1-n_j)$ is the hole number operator.  Different types of boundary conditions will be considered. 

This Hamiltonian is a modification of a model introduced by Fendley and Schoutens\cite{Fendley2006,Feher2017}, which we will refer to as $H(\vec1)$ because it is equal to the above Hamiltonian with all $\lambda_i=1$. This $H(\vec1)$ is a supersymmetric (SUSY) lattice model whose spectrum is completely solvable by nested Bethe ansatz. The energy levels of this strongly-coupled model are the same as those of the free fermion model, but with different degeneracies. Two particles, one on each chain, can form a coupled zero energy state, aptly named a ``Cooper pair''. When considering all possible particle numbers, there is a huge degeneracy of $2^{L/2}$ zero energy states due to these bound pairs. Any SUSY conserving modifications of the couplings leave these ground states intact, because the degeneracy is protected by the Witten index\cite{Witten1982}. In this paper we are interested in the behaviour of these bound pairs in a quantum pump context. 

There are two main ingredients added to the original model to obtain \eref{hamiltonian}. Firstly, the factors $\lambda_j$ are introduced in order to periodically modulate the couplings between the sites, see \ref{SUSY} for details. The specific method ensures that the new Hamiltonian \eref{hamiltonian} is still supersymmetric, so the number of $H(\vec1)$ ground states is still protected by the Witten index, but the wave functions get a $\lambda_i$ dependence. The values of $\lambda_j$ are chosen periodically such that $\lambda_{j+q} = \lambda_j$, where $q$ is the number of sites in a unit cell. So there are $q$ independent unique $\lambda_j$ values, which can be written as a $q$-dimensional vector $\vec\lambda = (\lambda_1,\ldots,\lambda_q)$. The original model is regained when setting $q=1$. The situation with $q=2$ will be considered in an upcoming article. For now we will concern ourselves with $q\geq3$, where the focus will be on $q=3$. All $\lambda_j\in\mathbb{R}$ have values in the interval $[0,1]$ and we add the extra restriction $\sum_i \lambda_i^2=1$ such that $\vec\lambda$ lives on part of a $q$-dimensional sphere $S^q$ with all $\lambda_j \geq 0$. On this sphere is one special point where all the $\lambda_j$ become equal to $1/\sqrt{q}$, which will be called $\vec\lambda_c$. At this point the Hamiltonian is reduced to $H(\vec1)$, but with an extra factor of $1/q$ in front. We will later show that by adiabatically transporting $\vec{\lambda}$ around $\vec{\lambda}_c$ a topological quantum pump will be created.

Secondly, the original $H(\vec1)$ was defined on an open chain while in this paper periodic boundary conditions will be considered in order to investigate the quantum pump using standard techniques. The Hamiltonian given above in \eref{hamiltonian} has periodic boundary conditions by simply setting $c_{j+L}=c_j$. However, this (weakly) breaks the SUSY. To keep the SUSY it is necessary to add a particle number dependent phase to the hops over the edge in both the upper and lower leg, see \ref{SUSY} for details. The SUSY periodic version of our Hamiltonian, which we will use for the calculations of this paper, is
\begin{equation}
\eqalign{K = \sum_{j=1}^{L-2}\lambda_{j+1}\,\lambda_{j}\, c_{j+2}^{\dagger} \,(p_{j+1}-in_{j+1})\,c_{j} + \rm{h.c.}\\
	\qquad + e^{i \chi_1} \lambda_L\,\lambda_{L-1}\,c^{\dagger}_1\,(p_{L}-in_{L})\,c_{L-1}+ \rm{h.c.} \\
	\qquad - e^{i \chi_2}\,\lambda_1 \, \lambda_L \,c^{\dagger}_2\,(p_{1}-in_{1})\,c_{L} + \rm{h.c.}\\
	V = \sum_{j=1}^{L}(\lambda_{j-1}^2 \,p_{j-1} + \lambda_{j}^2 \,p_{j+1})\,n_j} \label{hamiltonianPeriodic}  
\end{equation}
with
\begin{equation}
\chi_1 = {\pi \over 2} N_2 - \pi N_1 - \varphi (N_1+N_2), \quad \chi_2 = - {\pi \over 2} N_1 - \varphi (N_1+N_2) \ .
\end{equation}
Here $N_1=\sum_{i=1}^{L/2}n_{2i-1}$ and $N_2=\sum_{i=1}^{L/2}n_{2i}$ are the number of particles on the odd and even sites (or equivalently the lower and upper legs of the chain) and $\varphi$ is an angle that can be chosen freely. Equally dividing the phases $\chi_1$ and $\chi_2$ over all $L/2$ bonds on both chains leads to a SUSY formulation of the model with translational invariance and therefore momenta as good quantum numbers.

\section{Eigenstates of the periodic model}\label{sect:eigenstates}
For systems with $q\geq3$ the possible eigenstates and their energies are easy to understand when considering the situation with only one non-zero $\lambda_j$. Because of the translational symmetry the choice of which $\lambda_j$ is non-zero is arbitrary. This is not the case in the open chain, where the translational symmetry is broken at the edges. Let us take $\lambda_{1} = 1$ and $\lambda_{j=2,\ldots,q}=0$ to describe the possible eigenstates and their energies. There are two different types of sites in a unit cell. Sites 1 and 2 are a ``dimer'' with a finite potential energy and all other $q-2$ sites have a zero energy potential and are completely disconnected from the rest of the chain. The hopping term of \eref{hamiltonianPeriodic} can be disregarded, because there are never two consecutive $\lambda_j \neq0$.

With only one particle on the chain, the system has $E=1$ when this particle is placed on site 1 or 2 of a unit cell and $E=0$ when placed on one of the other sites. There will be three different energy levels when adding a second particle. If both particles occupy a dimer site in two different unit cells, then the system will have $E=2$. When one is on a disconnected site and the other on a dimer site, irrespective of the unit cells, it has $E=1$. And if both particles are on the disconnected sites (again irrespective of the unit cells), the system will have $E=0$. In addition, the energy will also be zero when the two particles form a coupled pair by occupying the two dimer sites in the same unit cell, due to the interaction term in the potential. Because all sites are either dimers or disconnected sites, no new possibilities will arise when adding even more particles and thus this general structure of eigenstates remains valid for all particle number sectors and all system sizes. The only way to add extra particles without adding extra energy is to either put a bound pair on a dimer or a single particle on a disconnected site. In this way it is possible for every value of $N$ to create at least one zero energy state. Note that it is not possible for every value of ($N_1$, $N_2$), where $N_1$ particles are placed on the upper leg and $N_2$ on the lower leg. From now on we will refer to these two different ground state contributions as ``bound pairs'' and ``single particles'' and leave it implicit that we're not referring to states with higher energies. 

This picture will hold in more general circumstances as long as only a maximum of two neighbouring $\lambda_j$ are non-zero and all other $\lambda_j$ are zero. It is thus possible to make a path in $\vec\lambda$-space where the ground states always remain at zero energy. This protocol will be called the ``control freak'' path, inspired by the ``control freak pump'' of the Rice-Mele model as described in \cite{Asboth2015}. In this protocol we start with $\lambda_1=1$ and all other $\lambda_j=0$. Then the value of $\lambda_1$ is lowered and $\lambda_2$ is raised until $\lambda_1=0$ and $\lambda_2=1$. Subsequently, $\lambda_2$ is lowered and $\lambda_3$ is raised until $\lambda_2=0$ and $\lambda_3=1$. This process is continued with the other $\lambda_j$ until the system returns to the original situation with $\lambda_1=1$ and all other $\lambda_j=0$. 

When more than two consecutive $\lambda_j$ are non-zero, the control freak energy levels (all at $E=n$ for integers $n\geq0$) will split into bands with states close to the original energy value. With Hamiltonian \eref{hamiltonianPeriodic} all states have $E\geq0$. With \eref{hamiltonian} some states will get negative energy, but all energies lie close to the energies of \eref{hamiltonianPeriodic} except for $\vec{\lambda}$ in the vicinity of $\vec{\lambda}_c$. Different values of $\varphi$ also change the energies by a small amount, especially close to $\lambda_c$, but it will never lead to the band gap closing. 

For general $\vec{\lambda}$ the localized eigenstates along the control freak path discussed above will spread out due to the hopping term. Because the Hamiltonian has cotranslation symmetry, it is possible to use the many body momentum basis states
\begin{equation}\label{DefKbasis}
	\left|k,n \right\rangle = \frac{1}{\sqrt{M}}\sum_{j=0}^{M-1} e^{ikj}T^j\left| n\right\rangle \\ 
\end{equation}
with $k = 2\pi l/M$ for $l\in(0,1,\ldots,M)$. Here $\left| n\right\rangle$ are the usual occupation basis states and $T$ is a cotranslational operator\cite{Qin2018} whose precise action depends on the state on which it acts. If $q$ is even, $T$ translates all particles over $q$ sites, i.e. it translates the full state by one unit cell. If $q$ is odd the action of the translation operator depends on the particle numbers per leg $(N_1,N_2)$. If $T$ translates the state by $q$ odd sites, $N_1$ and $N_2$ will be switched. Since these particle numbers should be conserved, this operation is only allowed for the particle number sectors with $N_1=N_2$. In order to conserve the particle numbers on each leg when $N_1\neq N_2$, $T$ should translate the lattice with $2q$ sites. The number $M$ is the minimal amount of times you need to apply $T$ to a basisstate to return to the original state. It is often equal to $p$ (or $p/2$ if $T$ translates over $2q$ sites), but this is not necessarily true for $\left| n\right\rangle$ with a repeating pattern of occupied and unoccupied states. With the basis states \eref{DefKbasis} the Hamiltonian \eref{hamiltonianPeriodic} can be block diagonalized as $H(k)$. The many body eigenstates of the $m$'th Bloch band $\left| \psi_m(k)\right\rangle = \sum_n \psi_m(k,n)\left| k,n\right\rangle$ can then be found by diagonalizing $H(k)$.
 
It is also possible to create the many body Wannier states \cite{Ke2017}
\begin{equation}\label{Wannier}
\left|W_m(R) \right\rangle = \frac{1}{\sqrt{p}}\sum_{k} e^{-ikR}\left|\psi_m(k)\right\rangle.
\end{equation}
Here $R$ is the site around which the Wannier state is centred.
Because there is full freedom in choosing the phase of each state $\left| \psi_m(k)\right\rangle$, it is possible to choose these phases in such a way to obtain maximally localized Wannier states\cite{Ke2017}. Therefore, even for general $\vec\lambda$ it is meaningful to talk about local bound pairs.

Because of the interaction term a many body state for general values of $\vec\lambda$ is not simply an addition of many single particle and coupled pair states, in contrast to states along the control freak path. Nevertheless, the lowest energy levels and eigenstates of the many body wave functions are observed to be close to those of a simple addition of single particles and bound pairs. This is due to the interaction only being local. It therefore makes sense to consider only the single particles and bound pairs when looking at pump behaviour. Later on we will numerically show that the pump with an arbitrary amount of particles works identical to that of many bound pairs and single particles, such that this assumption is indeed correct.

We split the lowest energy level in two levels, one with single particles and the other with bound pairs, by adding an extra onsite potential term 
\begin{equation}
H_w = \sum_{j=1}^{L}w(1-\lambda_{j-1})(1-\lambda_j)n_j 
\end{equation}
with $w$ some real constant. This term primarily shifts the energies of the single particle states. This can best be seen in the limit with $\lambda_1=1$ and $\lambda_{j\neq1}=0$. Because the bound pairs are next to $\lambda_1$, $H_w$ adds no extra energy to these states. In contrast, a single particle at site $j$ lies next to $\lambda_{j-1}=\lambda_j=0$, such that this state gets an extra energy of $w$. This $H_w$ term will therefore split the zero energy level in a level with all single particle states and a level with all the bound pairs. For $w>0$, the bound pairs will have the lowest energy. There is one bound pair per unit cell as long as $\vec\lambda\neq\vec{\lambda}_c$, so in the particle number sector $N=(p,p)$ there will be a non-degenerate ground state with all pairs occupied. When $w$ is negative, the single particles get the lowest energy. Now there will be a unique ground state when all these single particle states are occupied, which happens when both legs have $p(q-2)/2$ particles. This splitting of the ground state level allows us to investigate the pump behaviour of the single particles and bound pairs separately. Moreover, by occupying all single particles or bound pairs we have created a non-degenerate many body ground state with an energy gap to the excited levels, which is a requirement to get a topological quantum pump of the many body states. This $H_w$ does weakly break the SUSY, which primarily results in the original $E=0$ states obtaining a non-zero energy for general $\vec{\lambda}$.

\begin{figure}
	\centering
	\includegraphics[width=.65\linewidth]{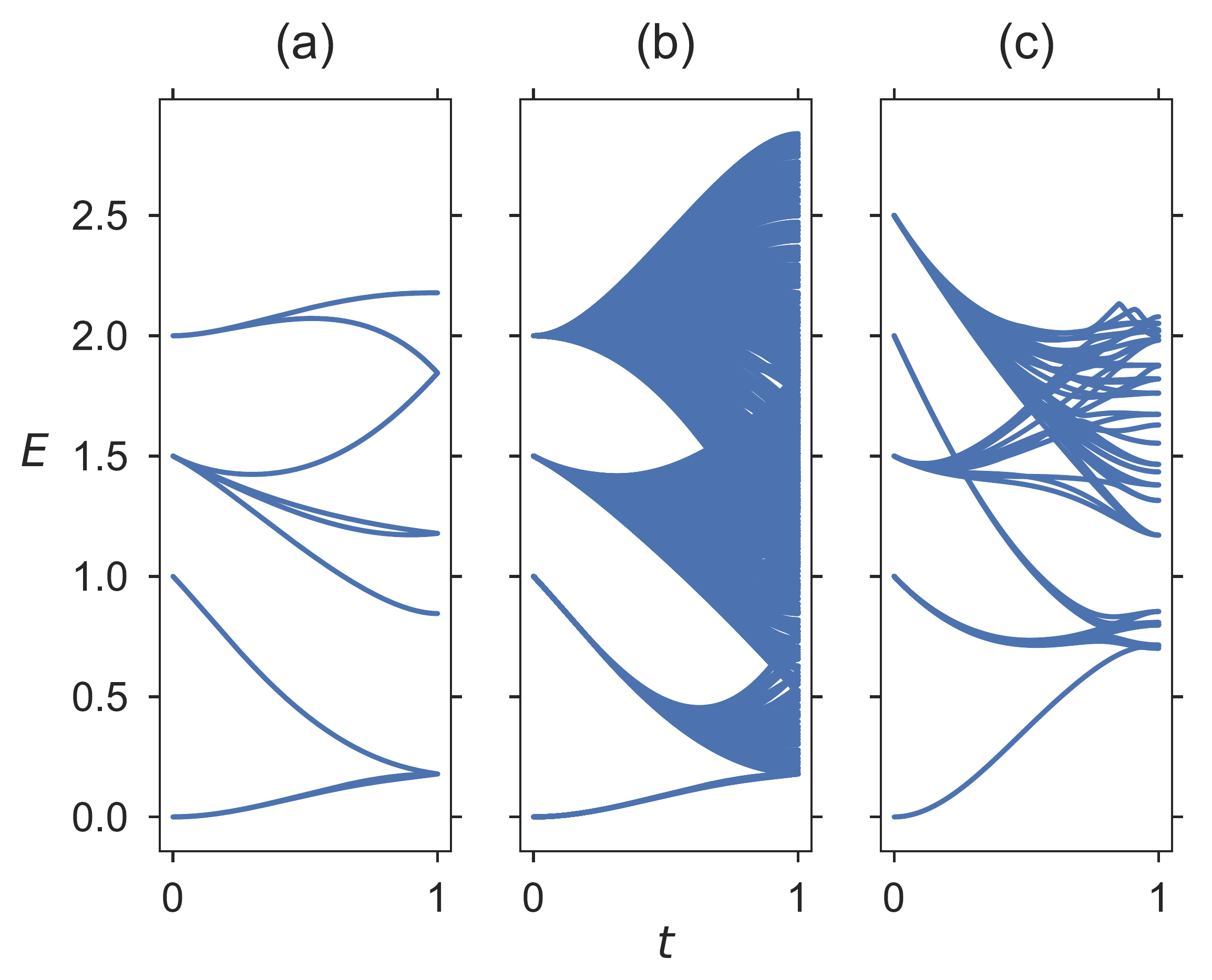}
	\caption{Eigenvalues along the path from $\vec\lambda(t=0)=(1,0,0)$ to $\vec{\lambda}(t=1)=\vec{\lambda}_c$ with $w=1/2$, $q=3$, and $\varphi=0$ for (a) $L=6$, $N=(1,1)$; (b) $L=72$, $N=(1,1)$; (c) 80 lowest eigenvalues for $L=12, N=(4,4)$.}
	\label{fig:eigenvaluesPairs}
\end{figure}
As an explicit example, see figure \ref{fig:eigenvaluesPairs} for the eigenvalues of the system with $q=3$, $w=1/2$, and $\varphi=0$ along the line $\lambda_1(t) = \sqrt{(1-2t^2/3)}, \lambda_2(t)=\lambda_3(t)=t/\sqrt{3}$, which interpolates from $\vec{\lambda}(t=0) = (1,0,0)$ with the localized control freak eigenstates described above to $\vec{\lambda}(t=1)=\vec\lambda_c=(1/\sqrt{3},1/\sqrt{3},1/\sqrt{3})$ which is the uniform model. The eigenvalues are shown for different particle numbers and system sizes. The left and central subfigures are for two systems with one particle on each leg. The left subfigure is for the smallest possible system of $L=6$ (two unit cells) and the central subfigure for the larger system of $L=72$. In both cases, the lowest band consists of all bound pairs (one pair for each unit cell) and is almost degenerate, meaning that the band structure is nearly flat. The band gap to the first excited band (consisting of states with two single particles with $E(t=0)=2w$) only closes at $t=1$ at the critical point. Comparing these two subfigures, it is clear that larger system sizes do not lead to band gap closing. In the right subfigure, with $L=12$ and $N_1=N_2=4$, the ground state is non-degenerate and it is the state with all two particle bound pairs occupied. Here too the band gap remains open for all $\vec{\lambda}\neq \vec{\lambda}_c$, showing that adding more particles does not introduce more band gap closing points. This band gap closing at $\vec{\lambda}_c$ can be explained as follows. At $\vec{\lambda}_c$ all $\lambda_i$ are equal, so the model has effectively become the $q=1$ model that can host one bound pair for each 2 sites, which is more than the original one bound state per $q$ sites. By symmetry all these bound states should have the same energy, meaning that the band gap between these levels should close at this point.

\begin{figure}
	\centering
	\includegraphics[width=.65\linewidth]{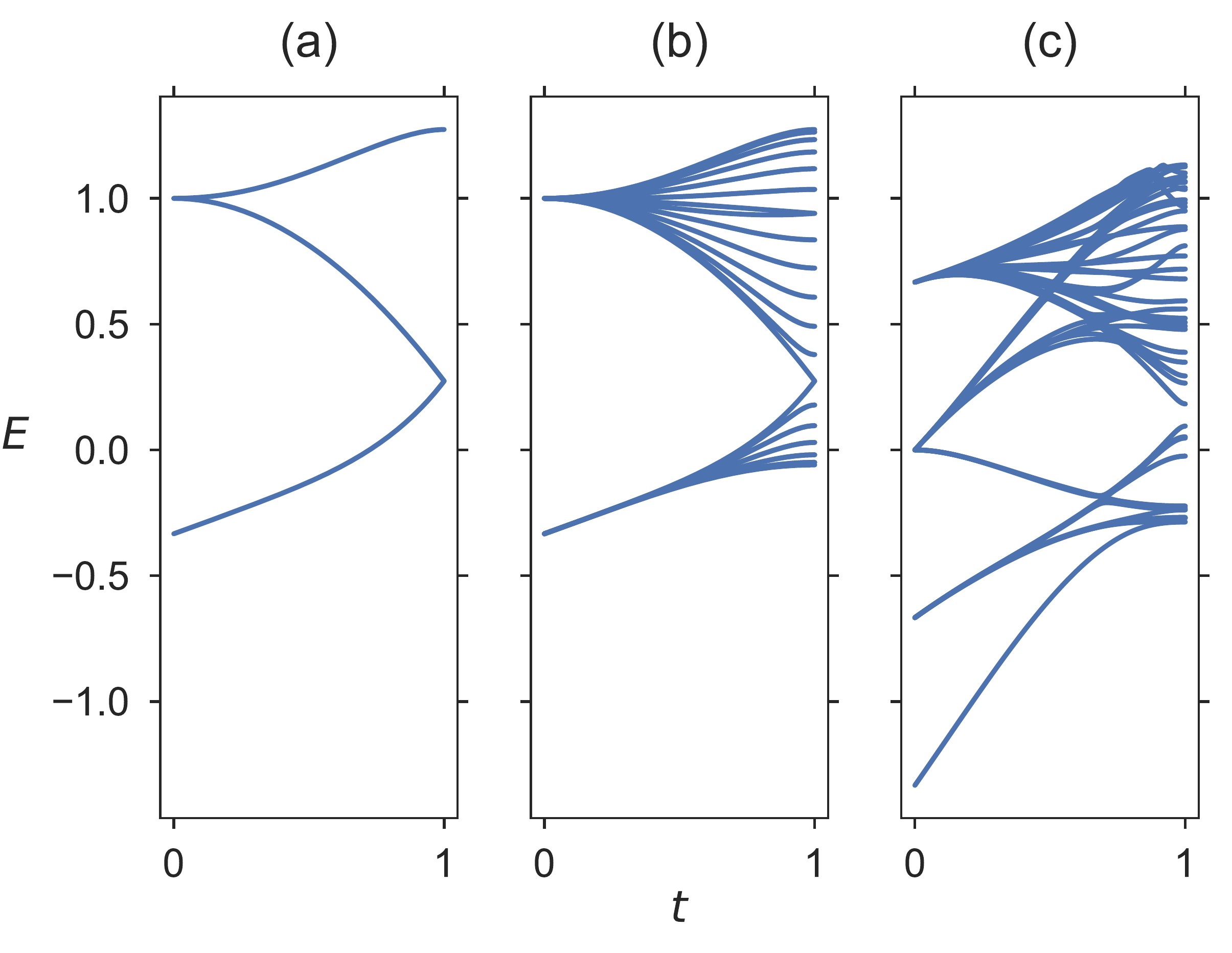}
	\caption{Eigenvalues along the path from $\vec{\lambda}(t=0)=(1,0,0)$ to $\vec{\lambda}(t=1)=\vec{\lambda}_c$ with $w=-1/3$, $q=3$, and $\varphi=0$ for (a) $L=6$, $N=(1,0)$; (b) $L=72$, $N=(1,0)$; (c) 80 lowest eigenvalues of $L=12, N=(2,2)$.}
	\label{fig:eigenvaluesFree}
\end{figure}

The same analysis can be done for the single particle states. In \fref{fig:eigenvaluesFree} the eigenvalues are plotted along the same path, but now with $w=-1/3$. The left and central subfigures are respectively for the systems with $L=6$ and $L=72$, both with one particle on the lower leg and no particles on the upper leg. The right subfigure is for $L=12$ and $N_1=N_2=2$. The lowest energy levels are now single particle states, there is one state per leg for every two unit cells (6 sites). In the right subfigure all single particle states are occupied, such that the lowest level is non-degenerate. Once again it is clear that the band gap between the two lowest levels only closes at $\vec{\lambda}=\vec{\lambda}_c$ for all three cases.

\section{Topological quantum pump}
\begin{figure}[h]
	\centering
	\includegraphics[width=0.45\linewidth, trim={0 4.5cm 0 0}, clip]{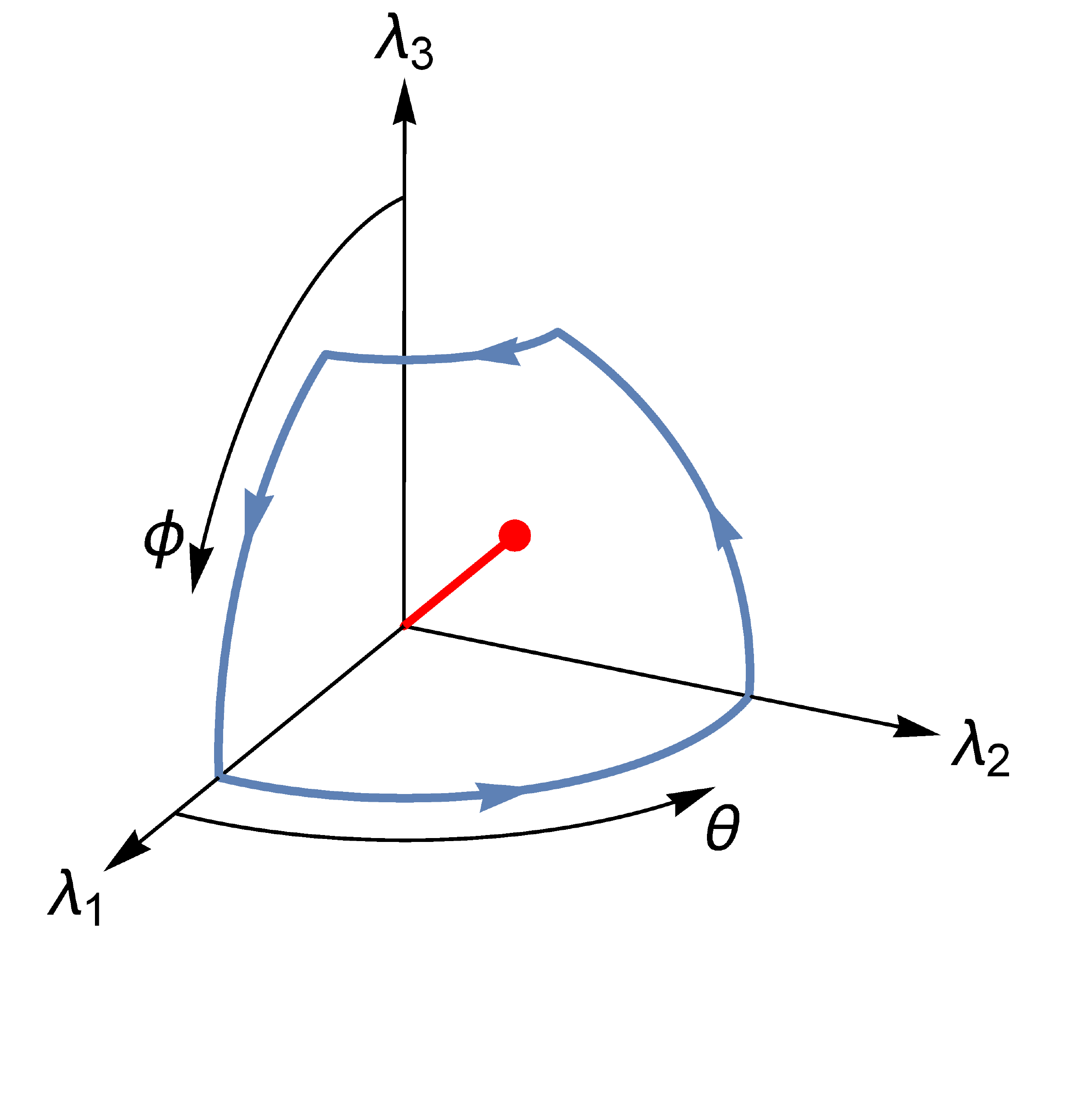}
	\caption{The $\vec{\lambda}$-space for $q=3$ and the path used for the calculations with $\epsilon=\pi/7$ (non-trivial). The red dot indicates the critical point.}
	\label{fig:pumppath}
\end{figure}

In this section we will consider the topology of the lowest energy bands. For $w>0$ this band consists of bound pairs and for $w<0$ of single particles. From now on we will focus on systems with $q=3$ and only remark on systems with a higher $q$ in the end, because the parameter space of $q\geq4$ is a lot richer compared to $q=3$, resulting in a more difficult situation. We will adiabatically drive the Hamiltonian $H(k,t)$ by making a closed loop in $\vec{\lambda}$-space with period $T$, i.e. $\vec{\lambda}(t+T)=\vec{\lambda}(t)$, such that $H(k,t+T)=H(k,t)$. As shown above, the critical point $\vec{\lambda}_c$ is the only point in $\vec{\lambda}$-space where the band gap between the lowest two bands closes. Specifically for $q=3$, $\vec{\lambda} = (\lambda_1,\lambda_2, \lambda_3)$ lives on the part of a three dimensional sphere of radius 1 with all $\lambda_j$ values positive, see \fref{fig:pumppath}. The critical point is located at $\lambda_1=\lambda_2=\lambda_3=1/\sqrt{3}$. Consequently, two families of topologically different closed paths $\vec{\lambda}(t)$ can be distinguished where the band gap always remains open along the path. A topologically trivial path is one where the critical point lies outside the loop. Such a path is topologically equivalent to not changing $\vec{\lambda}$ at all. A closed path with the critical point inside the loop is topologically distinct. We will show that along a topologically trivial path no charge gets pumped and along a non-trivial path a charge is transported through the chain. In the limit of infinite length and pump time, this charge is exactly quantized and is equal to a non-zero (many body) Chern number.

The Chern number is defined as
\begin{equation}\label{Chern}
C_m = \frac{i}{2\pi}\int_0^{2\pi}dk\int_{0}^T dt  \left(\left\langle\partial_t\psi | \partial_k\psi \right\rangle - \left\langle\partial_k\psi | \partial_t\psi \right\rangle\right).
\end{equation}
Here $\left| \psi\right\rangle= \left| \psi_m(k,t)\right\rangle = \sum_n \psi_m(k,t,n)\left| k,n\right\rangle$ are the eigenstates of the $m$'th Bloch band of $H(k,t)$. This definition can be used for both single particle Chern numbers and many body Chern numbers by choosing the appropriate basis states $\left|n\right\rangle$.

There are two different ways in which the pump behaviour can be observed and the corresponding Chern numbers can be calculated. Firstly, we will consider the $(1,1)$ particle number sector for the bound pairs and the $(1,0)$ sector for the single particles. These sectors have at every time $t$ a (many body) Bloch band as a function of momentum $k$. To this Bloch band a localized Wannier state $\left| W(t,R)\right\rangle$ \eref{Wannier} can be associated. Following \cite{Thouless1983, Ke2017}, the (many body) Chern number \eref{Chern} can be related to the transport of the Wannier state as
\begin{equation}\label{chernWannier}
C_m= \Delta P /q,
\end{equation}
 where $m$ is the band number, $q$ the number of sites in a unit cell, and $\Delta P$ is the number of lattice sites in which the center of mass $\left<\hat{x}(t)\right>$ (with $\hat{x}$ having units of the lattice site number) of a Wannier state is transported after one pump cycle. By starting out in a localized Wannier state at $t=0$ and following the center of mass during time evolution, the Chern number can be measured. It should be noted that the definition of $\hat{x}$ is inherently vague in the the case of periodic boundary conditions, because sites 1 and $L$ are next to each other. This method is therefore more accurate if the state is centred around site $L/2$. Consequently, this method can not be used to calculate the displacement of the fully occupied bands, because of the uniform occupation.

Secondly, it is also possible to measure the topology with the particle number flux through an arbitrary cross section of the chain \cite{Thouless1983, Asboth2015}. The particle number flux through a segment $S$ of the chain can be measured with the operator $J_S(t) = -i \left[n_S,H\right]$, here $n_S$ is the number operator over all sites in the segment $S$, i.e. $n_S=\sum_{i\in S}n_i$. This $J_S$ can then be split in $J_L$ and $J_R$, which are the number flux operators through the left edge and the right edge of the segment. The $J_L$ and $J_R$ turn out to be equal to the matrix elements of $K(t)$ for the states which hop over the corresponding edges. The charge pumped through one of those edges during a pump cycle is equal to 
\begin{equation}\label{chargeInt}
	Q = \int_0^T \left\langle \psi(t)| J_{L/R}(t)|\psi(t) \right\rangle dt.
\end{equation}
 Here $\left| \psi(t)\right\rangle$ is not the instantaneous wave function at time $t$, but rather the state obtained by adiabatically time evolving the starting state $\left| \psi(0)\right\rangle$. Note that the direction of the measured current is different for the left and right edge, which should be taken into account when interpreting the result. For a fully occupied non-degenerate ground state level of a non-interacting or weakly interacting (many body) Hamiltonian, $Q$ is equal to the (many body) Chern number for an infinite chain under adiabatic conditions\cite{Thouless1983,Niu1984}. It is not clear that in the case of Hamiltonians with strong interactions this $Q$ will be equal to the many body Chern number associated to the bound pairs. However, in the next section we will show numerically that in our model this equality will hold. This method is useful to calculate the Chern number of the many body ground state of fully occupied bound pairs or single particles. It can also be used for the Wannier states by considering a site next to the starting point of the Wannier state. However, due to the spread out nature of these states for general $\vec{\lambda}$, this method is not perfect and the center of mass method is preferred. The computation time grows exponentially with the particle number and from experience only small systems (up to $L=12$) with fully occupied lowest energy bands can be calculated within a reasonable time under adiabatic conditions. For the (1,1) sector much larger systems can be accurately calculated with the center of mass method.

\section{Results}\label{results}
The following method was used to obtain the results. We used $q=3$, so as noted before $\vec{\lambda}$ lives on the part of a three dimensional sphere of radius 1 with all $\lambda_j$ values positive. The critical point is located at $\lambda_1=\lambda_2=\lambda_3=1/\sqrt{3}$. The standard spherical coordinates $(\theta, \phi)$ are used to parametrize these $\lambda_j$ as $\lambda_1 = \cos(\theta)\sin(\phi)$, $\lambda_2 = \sin(\theta)\sin(\phi)$, and $\lambda_3 = \cos(\phi)$, see \fref{fig:pumppath}. For all paths we start out with $\lambda_1 = 1$ and $\lambda_2=\lambda_3=0$, i.e. $\theta=0$ and $\phi=\pi/2$. Then we do the following four steps to make a loop in the parameter space. In step one change $\theta = 0 \rightarrow \theta=\pi/2$, then in the second step $\phi= \pi/2 \rightarrow \phi=\epsilon$, in step three $\theta= \pi/2 \rightarrow \theta=0$, and finally in step four $\phi= \epsilon \rightarrow \phi=\pi/2$. The parameter $\epsilon$ is used to interpolate between topologically trivial and non-trivial paths. The critical point is inside the loop for $\epsilon < \arccos\left(1/\sqrt{3}\right)\approx0.30\pi$. As an example, the path with $\epsilon = \pi/7$ is shown in \fref{fig:pumppath}. The advantage of these specific paths is that the states at the starting point are exactly known and especially the maximally localized two body Wannier states are easy to construct. Explicitly, the state $c^\dagger_{3i+1}c^\dagger_{3i+2}\left| 0\right\rangle$ for $i \in \mathbb{N}$ is already a maximally localized Wannier state of a bound pair for $t=0$, because $\lambda_1=1$ and $\lambda_2=\lambda_3=0$, so any bound pair sits on the first two sites of a unit cell. Similarly, the states $c^\dagger_{3i}\left| 0\right\rangle$ are localized Wannier states for the single particles. We can thus choose the starting state $\left| \psi(t=0)\right\rangle$ and obtain the time evolved states $\left| \psi(t)\right\rangle$ by numerically integrating the Schr\"odinger equation with the time dependent Hamiltonian in which $\vec{\lambda}(t)$ follows the path described above. The total time of one pump cycle is $T$ and all four steps of the path have a duration of $T/4$. 

\begin{figure}
	\centering
	\includegraphics[width=0.65\linewidth]{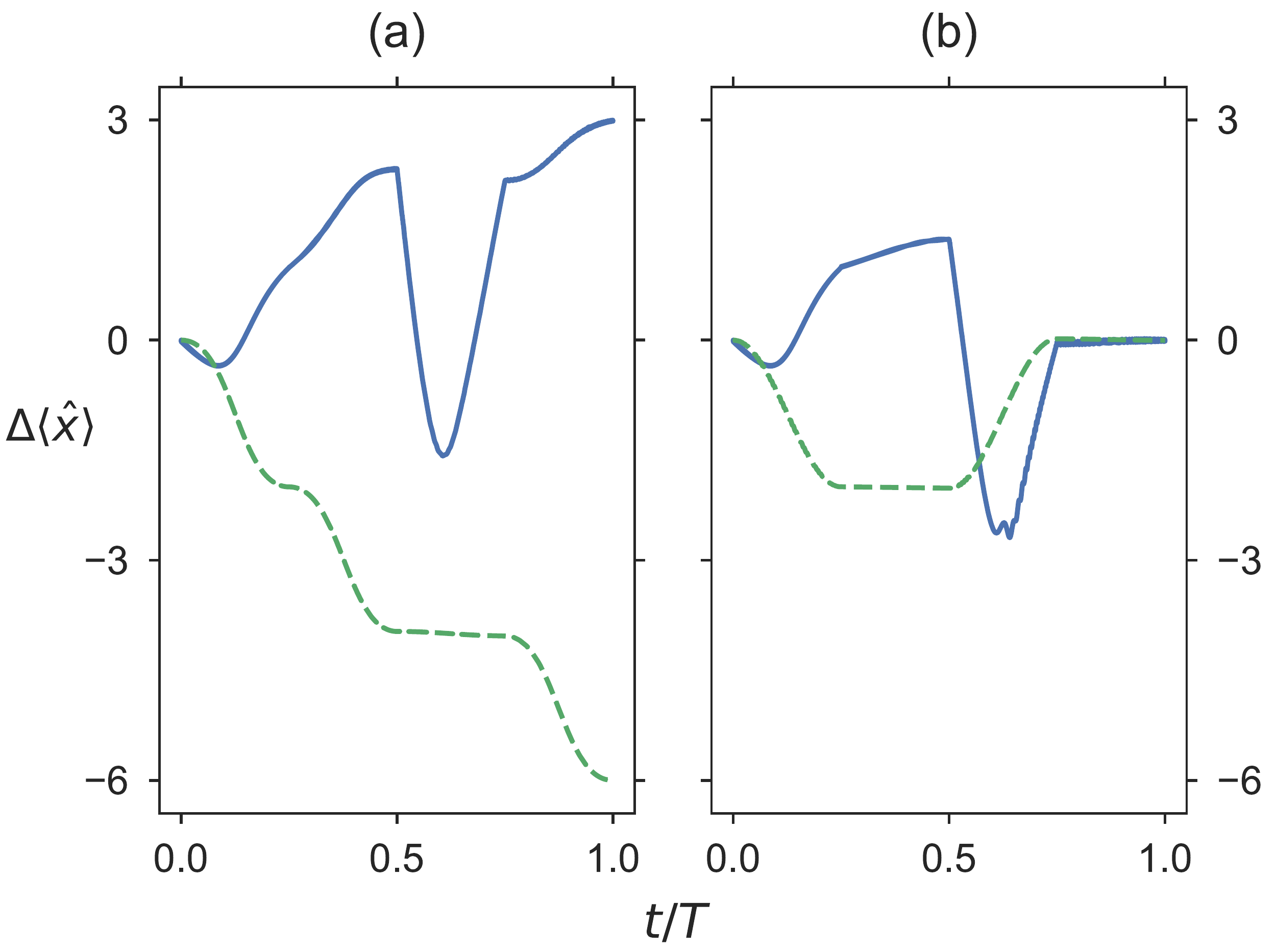}
	\caption{Displacement of the center of mass position during one pump cycle for a coupled pair (blue, solid) and a single particle (green, dashed) along (a) a non-trivial path and (b) a trivial path.}
	\label{fig:xmeanPumped}
\end{figure}

In \fref{fig:xmeanPumped} the shift of the center of mass position during the time evolution of a localized Wannier state on a chain with $L=48$ is shown for paths in the trivial and non-trivial regime. For the calculations of the bound pairs $T=5000$ was used along the path with $\epsilon=0.20\pi$ in the non-trivial regime and $\epsilon=0.40\pi$ in the trivial regime, both with $T=5000$ and $\varphi=0$. For the single particle state $\epsilon=0.05\pi$ was used for the non-trivial regime and $\epsilon=0.45\pi$ for the trivial regime, both with $T=1000$ and $\varphi=0$. The different values were chosen such that the paths are away from the control freak limit, but still avoid the effect of $\hat{x}$ being ill-defined between sites $L$ and 1, which mainly affected the single particle states. Additionally, for all calculations the starting states were close to the centre of the chain to minimize this problem. For the calculations involving the bound states we set $w=1/2$ and for the single particle states $w=0$. A non-zero $w$ is unnecessary in this case because there is only one particle, so no bound pairs with the same energy exist. In the trivial regime the total displacement $\Delta P = \left<\hat x (T)\right>-\left<\hat x (0)\right>\approx 0$ for both the single particle and the bound pair, so from \eref{chernWannier} it follows that the (many body) Chern number $C=0$ for both. In the non-trivial regime $\Delta P \approx 3$ for the bound pair, so $C=1$, and $\Delta P \approx -6$ for the single particle, so $C=-2$. 

\begin{figure}
	\centering
	\includegraphics[width=0.4\linewidth]{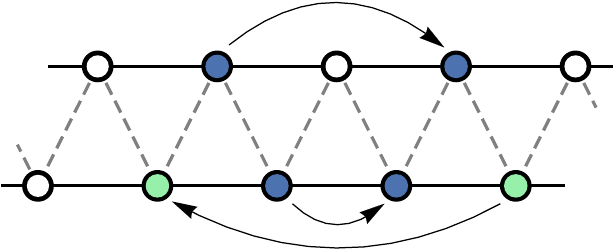}
	\caption{A coupled pair (dark blue) is pumped one unit cell (three sites) to the right and a single particle (light green) is pumped two unit cells (six sites) to the left.}
	\label{fig:chainHop}
\end{figure}
Along the pump freak path these results can easily be interpreted, see \fref{fig:chainHop}. Every bound pair moves one unit cell to the right during each pump cycle and the two particles hop over each other multiple times. The single particle moves in the opposite direction and moves twice as fast. We've also observed that in systems with both bound states and single particles, the two do not collide but move unaltered past each other, reminiscent of solitons.

\begin{figure}
	\centering
	\includegraphics[width=0.6\linewidth]{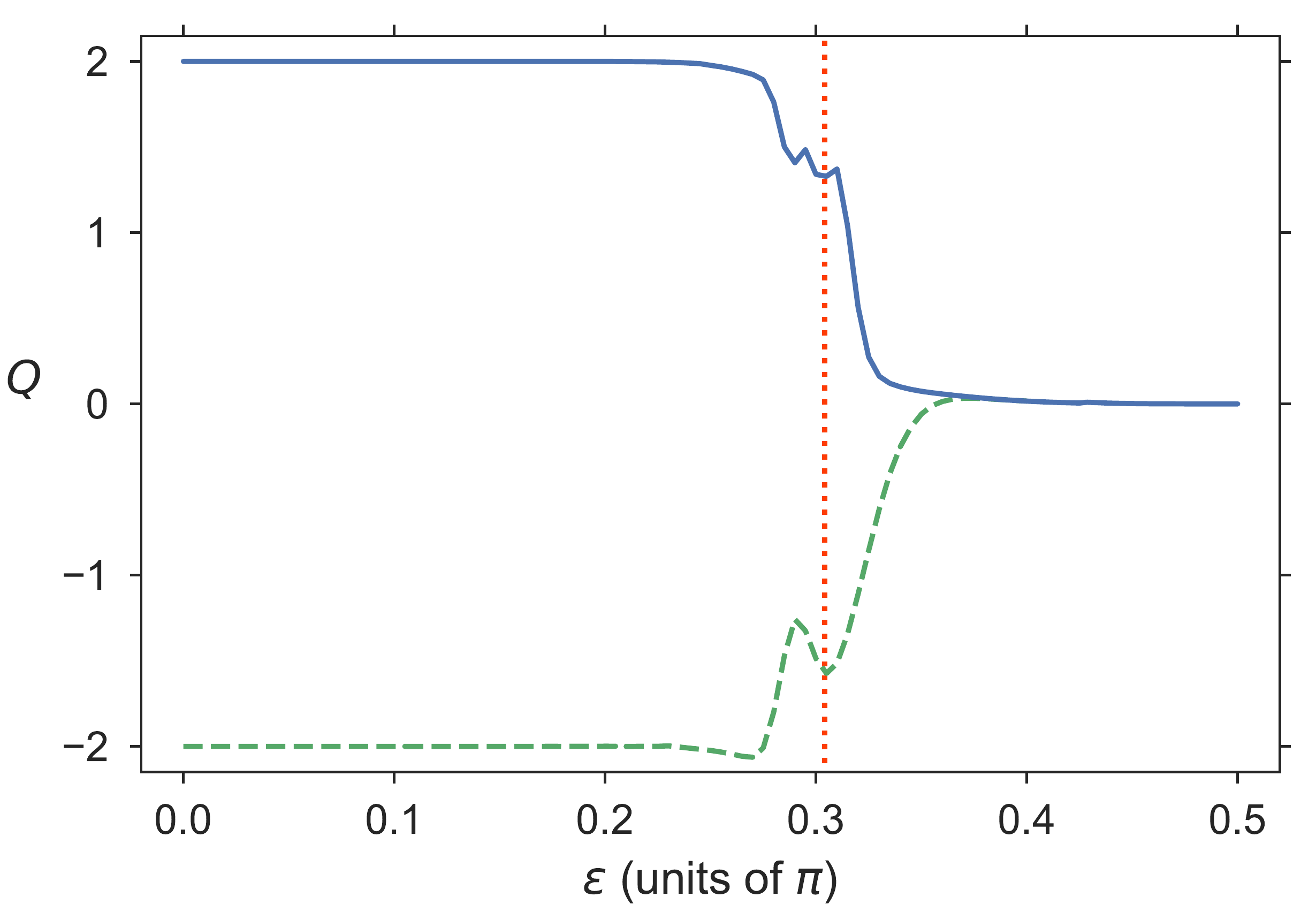}
	\caption{Total charge pumped during one cycle along paths with different $\epsilon$ for the lowest energy level with $L=6$, $T=10000$ for $0.25<\epsilon<0.43$ and $T=2000$ elsewhere. For the paired states (blue, solid) $N = (2,2)$, $w=1/2$, and for the single particle states (green, dashed)  $N = (1,1)$, $w=-1/3$. The dotted vertical line indicates the path with the critical point.}
	\label{fig:chargePumped}
\end{figure}
In \fref{fig:chargePumped} the total charge $Q$ \eref{chargeInt} pumped to the right through one point in the chain is shown for paths with different values of $\epsilon$. For the calculations of the fully occupied bound pair state $L=6$, $N=(2,2)$, $w=1/2$, and $\varphi=\pi/8$ were used. For the single particle states calculations $L=6$, $N=(1,1)$, $w=-1/3$, and $\varphi=0$ were used. To the left of the critical point are the topologically non-trivial paths and to the right the topologically trivial paths. As can be seen, the pumped charge tends to $Q=\pm2$ for a non-trivial path and $Q=0$ for a trivial path. The difference between the calculated values from the exactly quantized theoretical values, especially around $\vec\lambda_c$, comes from finite size effects\cite{Li2017} and imperfect adiabatic conditions. We have observed that for values of $\epsilon$ close to the critical point larger system sizes and larger values of $T$ improve the results, see also \ref{finitesize}. The adiabatic condition is $T\gg \Delta E$, where $\Delta E$ is the minimum of the band gap during the whole pump cycle. Since $\Delta E = 0$ at $\lambda_c$, this adiabatic condition will inherently be violated in calculations where the path is close to $\lambda_c$, because the pump has to be driven unrealistically slow. For the data in \fref{fig:chargePumped} the value $T=10000$ was used for paths with $0.25<\epsilon<0.43$ and $T=2000$ for all other paths.

These results of the pumped charge agree with the previous results of the displacement of the Wannier states. For the single particle state the displacement was $\Delta P=-6$, corresponding to a Chern number of $-2$. In a non-interactive charge pump with a fully occupied level this would lead to a quantized total pumped charge of $Q=-2$. In \fref{fig:chargePumped} it is shown that this result still holds true in our interactive model. Above, the many body Chern number of the bound pairs was found to be $C=1$. In \fref{fig:chargePumped} it can be seen that the total charge pumped in the fully occupied state is $Q=2$. Similar to the single particle states, these two results for bound pairs match, because every bound pair has a charge of $Q=2$. Therefore, the many body Chern number of the bound pairs in our system can be identified with the current pumped through a cut in the chain.

Most interestingly, the displacement calculations were done with one or two particles on a chain, while the charge calculations were done with many particles fully occupying the lowest energy band. This means that there are no extra interaction effects that play a role in determining the pump behaviour. The topological response of the full many body state can therefore be effectively described by that of independent single particles and bound pairs. For pump cycles along the control freak path this is easily understood because at all times the model is described by disconnected segments. Our results show that this is the case for any path through the parameter space. Even though small system sizes were used for the results shown above, we are confident that this result holds for larger system sizes. For states along the pump freak path this is clearly true, as explained in \sref{sect:eigenstates}, as well as for the states at $\vec\lambda_c$ which is the $H(\vec1)$-model\cite{Fendley2006}. In figures \ref{fig:eigenvaluesPairs} and \ref{fig:eigenvaluesFree} it was already shown that the energy gap for all $\vec{\lambda}\neq\vec{\lambda}_c$ remains open for larger system sizes. Additionally,  calculations performed on larger system sizes showed results closer to the quantized values compared to the small system size calculation from above, see \ref{finitesize}.

\section{Edge states in the open system}\label{openBoundary}
\begin{figure}
	\centering
	\includegraphics[width=0.6\linewidth]{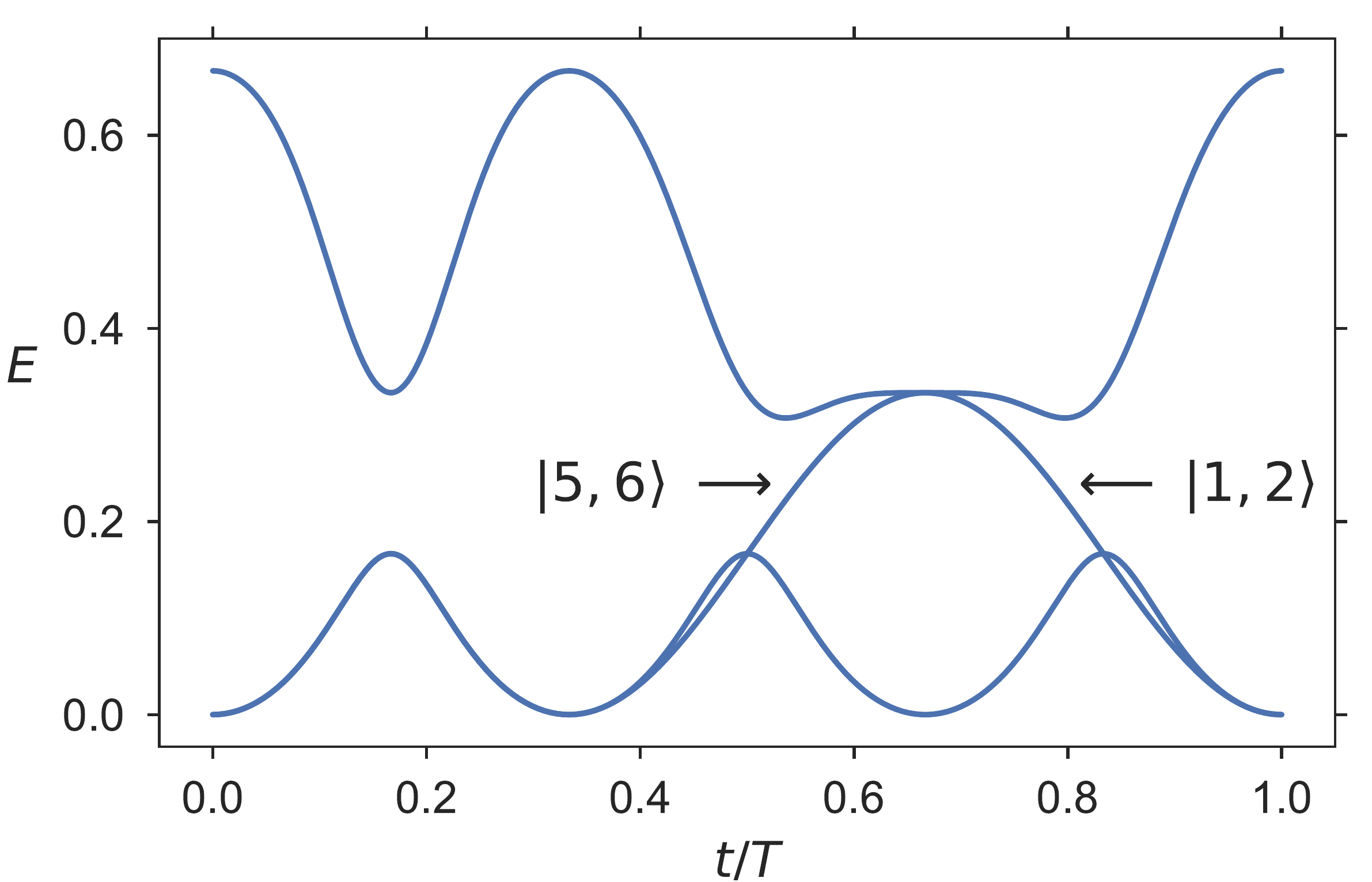}
	\caption{The three lowest energy levels of an open chain of $L=6$, $N=(1,1)$, and $w=1/3$ along the control freak path. The band gap crossing edge states are indicated.}
	\label{fig:eigenvaluesOpen}
\end{figure}
With open boundary conditions the energy spectrum gets band gap crossing edge states along topologically non-trivial paths. For the pump freak protocol the states can easily be tracked and the pump behaviour explained. See \fref{fig:eigenvaluesOpen} for the two lowest energy levels along the control freak path for $L=6$ with $q=3$ and $w=1/3$. We chose this small system size for simplicity of the energy levels, but the following explanation also holds for larger system sizes. Because of these boundary conditions $\lambda_L$, giving the strength between the first and last site, is zero, thereby breaking the symmetry between the different $\lambda_i$ components of $\vec{\lambda}$. For example, the system with $\lambda_1=1$ and the system with $\lambda_3=1$ will have qualitatively different eigenstates. 

\begin{figure}
	\centering
	\includegraphics[width=0.65\linewidth]{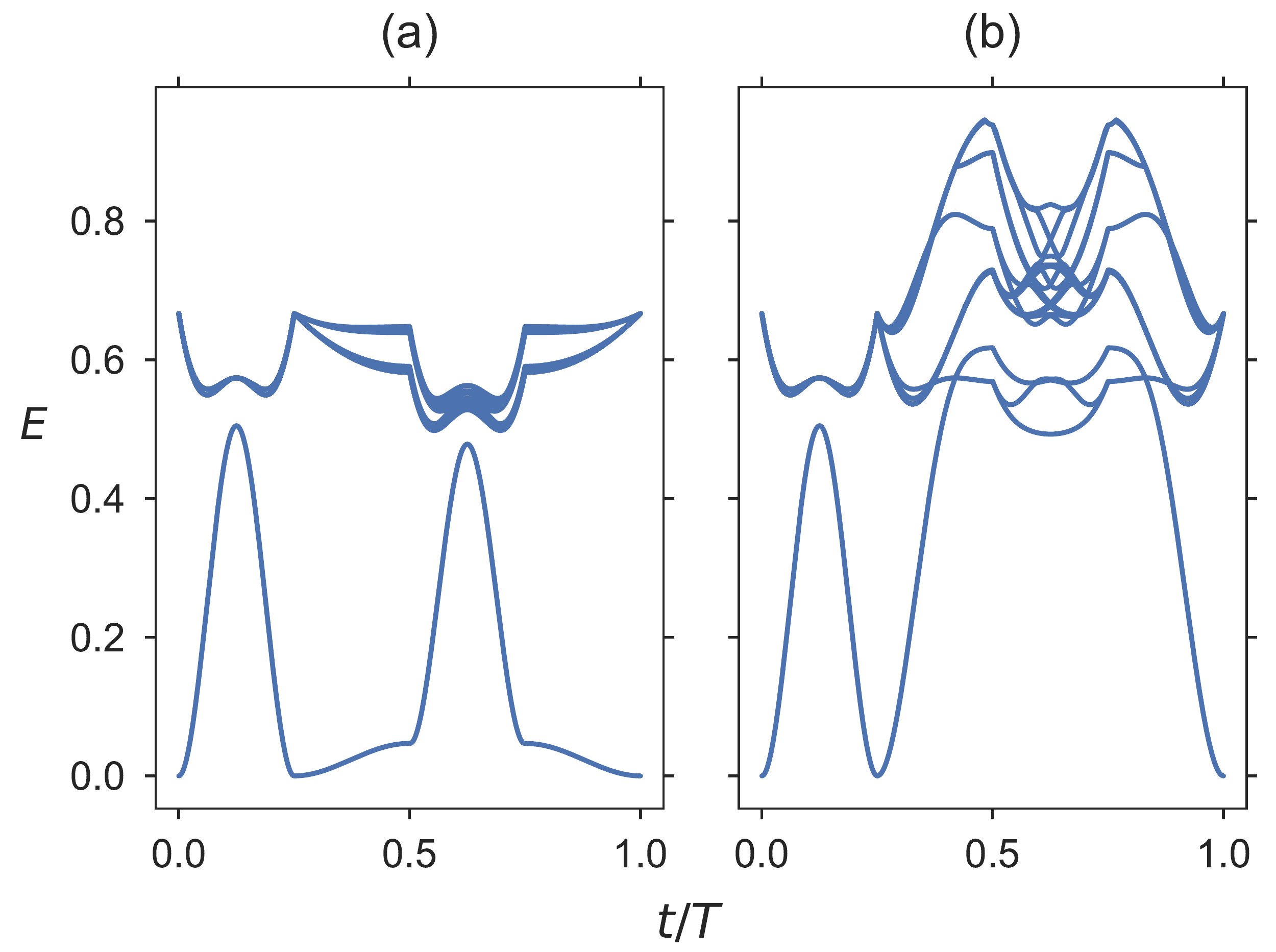}
	\caption{The two lowest energy bands of an open chain of $L=12$, $N=(4,4)$, and $w=1/3$ along the pump paths with (a) $\epsilon=0.45\pi$ and (b) $\epsilon=0.20\pi$.}
	\label{fig:eigenvaluesOpenMany}
\end{figure}
We choose to use the paths described before, i.e. starting with $\lambda_1=1$ and the corresponding eigenstates. The lowest level consists of the bound pairs with $E(t=0)=0$ and the first excited level of single particles with energy $E(t=0)=2w$. Along a topologically non-trivial path, local pairs are pumped to the right in agreement with the results shown in \sref{results} for the periodic chain. When they reach the right edge, they are separated and get pumped higher in energy to the single particle band. This is the band gap crossing edge state $\left| 5,6\right\rangle = c^{\dagger}_5c^{\dagger}_6\left| 0\right\rangle$ in \fref{fig:eigenvaluesOpen}. These single particles are pumped towards the left, where they recombine in a zero energy local pair at the left edge, which is the second band gap crossing edge state $\left| 1,2\right\rangle$. We checked that along topologically trivial paths the states return to their state at $t=0$ after each pump cycle, which is not the case for topologically non-trivial paths. This is for example clear when plotting the eigenvalues of the state with all bound pairs occupied, see  \fref{fig:eigenvaluesOpenMany}. Along a trivial path ($\epsilon=0.45\pi$ in the figure) this lowest non-degenerate state always has a band gap with the first excited state, therefore having to return to the beginstate after one pump cycle by the adiabatic theorem. For the topologically non-trivial path ($\epsilon=0.20\pi$) this band gap closes somewhere along the path such that the state gets pumped to a higher energy level.

\section{Generalization to larger unit cells}
\begin{figure}[!ht]
	\centering
	\includegraphics[width=0.5\linewidth]{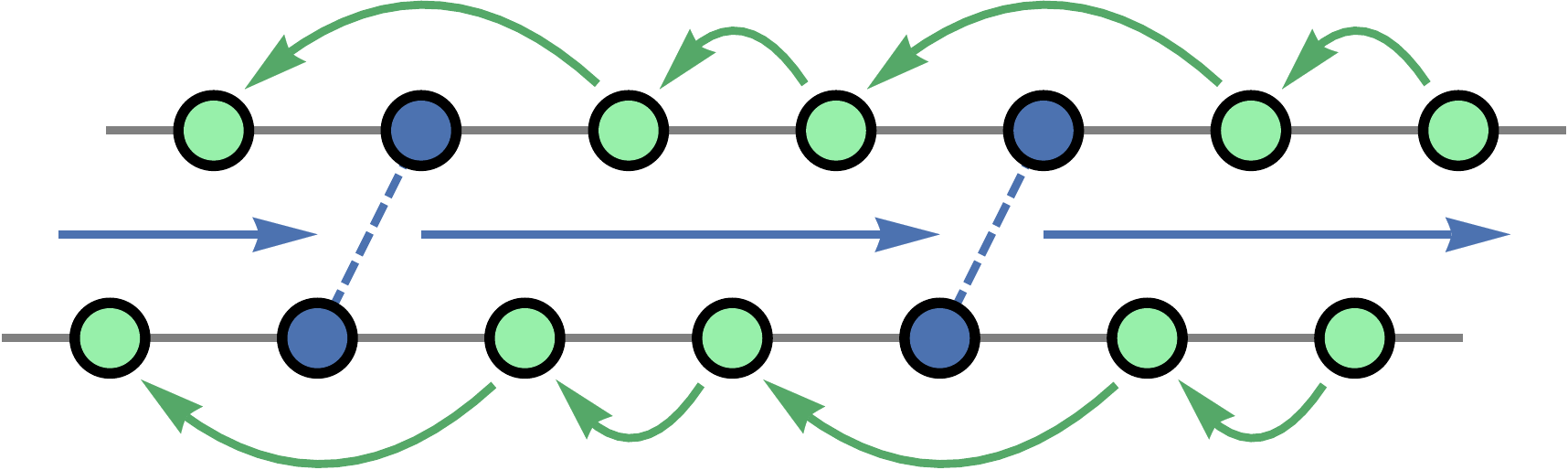}
	\caption{Particle transport during one pump cycle for $q=6$. The coupled pair (blue) is pumped one unit cell to the right. All single particles (light green) are pumped to the next available site to the left.}
	\label{fig:chainHopQ6}
\end{figure}

The model with $q=3$ is clear in terms of the topology. The $\vec\lambda$ can only be moved in a two dimensional space (parametrized by $\theta$ and $\phi$) and there is one clear critical point where all $\lambda_i$ become equal and the model effectively is described by the $q=1$ model. When generalizing to higher $q$ values, the situation becomes more difficult. For example, in the $\vec{\lambda}_{q=6}$ space there is a subspace where $\lambda_1=\lambda_2$, $\lambda_3=\lambda_4$ and $\lambda_5=\lambda_6$. This is effectively again the $q=3$ model (with all $\lambda_i$ values scaled by $1/\sqrt{2}$) which for example results in the possibility of two bound pairs per unit cell of $q=6$ sites. Within this subspace is also the critical point where all $\lambda_i=1/\sqrt{6}$. This makes the identification of different topologically (non-)trivial loops difficult. There is one situation that remains simple, which is the control freak path. For each unit cell there is one paired state that starts with energy 0 and $q-2$ single particle states with energy $w$. Then going along the path once, the paired state is pumped one complete unit cell to the right, just as in the $q=3$ model. All single particles move to the nearest available site to the left of them. In the case of $q=3$ this means they move two whole unit cells, for $q=4$ one unit cell and for $q\geq5$ they move either one site or two sites (jumping over the paired state), see \fref{fig:chainHopQ6}. The result is that the total pumped charge of the fully occupied pair and single particle levels is opposite of each other. During each pump cycle exactly $Q=2$ gets pumped to the right because of the bound pairs and $Q=2$ to the left because of the single particles.

\section{Summary}
We have introduced an interactive model of fermions on two coupled chains. The lowest energy levels of this model consist of single particles and/or coupled pairs based on the value of the parameter $w$. We have shown that by adiabatically changing the system parameters $\vec{\lambda}$, a topological quantum pump can be created. There is one critical point in the space of $\vec{\lambda}$ such that there can be topologically trivial and non-trivial loops in this space. No charge gets pumped when the path is trivial. During one pump cycle along a non-trivial loop, the bound pairs and single particles move in opposite directions. Specifically for $q=3$, the single particle Chern number is $C=-2$ and the many body Chern number of the bound pairs is $C=1$. This shows a novel way in which interactions lead to unexpected pump behaviour. It was also shown that in this model the particle current of the fully occupied bound pair states can be identified with the two body Chern number of the bound pairs.

\ack{KjS wishes to thank the Erwin Schr\"odinger Institute in Vienna, where part of this work was done, and to acknowledge illuminating discussions with Masaki Oshikawa. This work is part of the Delta ITP consortium, a program of the Netherlands Organisation for Scientific Research (NWO) that is funded by the Dutch Ministry of Education, Culture and Science (OCW).}

\appendix
\section{Supersymmetric boundary conditions}\label{SUSY}

The original Hamiltonian by Fendley and Schoutens\cite{Fendley2006} is constructed with the SUSY operators
\begin{equation}\label{DefQ}
Q = c_2^\dagger c_1 + \sum_{k=1}^{L/2-1} \left(e^{i\frac{\pi}{2}\alpha_{2k-1}} c^\dagger_{2k} + e^{i\frac{\pi}{2}\alpha_{2k}} c^\dagger_{2k+2}\right)c_{2k+1}
\end{equation}
where
\begin{equation}\label{DefAlpha}
\alpha_k = \sum_{j=1}^{k} (-1)^j n_j.
\end{equation}

The operator $Q$ brings particles from the odd leg to the even leg and $Q^\dagger$ does the reverse. It therefore changes the ``type'' of particle, just as in many other SUSY models fermions and bosons are related to each other by SUSY operators. The Hamiltonian is then constructed as $H=\{Q,Q^{\dagger}\}$, from which it follows that $[Q,H]=0$ and $[Q^\dagger,H]=0$. All solutions have $E\geq0$. States with $E>0$ form doublets (i.e. $\left| \phi\right\rangle$ and $Q\left| \phi\right\rangle$ have the same energy because $Q$ commutes with $H$) and $E=0$ ground states are annihilated by both $Q$ and $Q^\dagger$. These $E=0$ ground states are the bound pairs on which we focus in the main text.

To obtain the Hamiltonian \eref{hamiltonianPeriodic} we add two ingredients to \eref{DefQ}. The first addition is to implement the factors $\lambda_j$ by changing all $c^{\dagger}_{j+1}c_j\rightarrow \lambda_j c^{\dagger}_{j+1}c_j $ inside the definition of $Q$. This $\lambda_j$ therefore controls the strength between sites $j$ and $j+1$. This leads to the periodically varied Hamiltonian with open boundary conditions which is used in section \ref{openBoundary}.
 Secondly, equation \eref{DefQ} has open boundary conditions and we want to get a Hamiltonian with periodic boundary conditions. One requirement for the SUSY operators is that $Q^2=0$. We therefore add $- e^{i\chi} c^{\dagger}_L c_1$ to \eref{DefQ}, square it and impose that this expression should be 0 to find phases $\chi$ for which this new $Q$ is a SUSY operator. It turns out that this is the case for $\chi ={\pi \over 2} N_1 + \varphi (N_1+N_2)$ with $N_1$ the total number of particles on the odd sites and $N_2$ the number of particles on the even sites (or equivalently, the lower and upper chain), and $\varphi$ an angle that can be chosen freely. Our SUSY operator is then
\begin{equation}\label{DefQlab}
\eqalign{
	Q= \lambda_1 c_2^\dagger c_1 &+ \sum_{k=1}^{L/2-1} \left(\lambda_{2k}e^{i\frac{\pi}{2}\alpha_{2k-1}} c^\dagger_{2k} + \lambda_{2k+1}e^{i\frac{\pi}{2}\alpha_{2k}} c^\dagger_{2k+2}\right)c_{2k+1}\\
	& - \lambda_L e^{i{\pi \over 2} N_1 + i \varphi (N_1+N_2)} c^{\dagger}_L c_1}
\end{equation}
From this definition Hamiltonian \eref{hamiltonianPeriodic} is obtained by calculating $\{ Q , Q^{\dagger} \}$.

The Witten index for the model with periodic boundary conditions that preserve SUSY is the same as for open boundary conditions: $W=2^{L/2}$. The corresponding supersymmetric groundstates correspond to all possible combinations of local pairs, for which $L/2$ independent $E=0$ states are available \cite{Fendley2006}.

\section{Finite size effects and adiabatic conditions}\label{finitesize}
\begin{table}[h]
	\caption{\label{table1}Pumped charge $Q$ with $T=10000$ for different system sizes.}
	\begin{indented}
		\item[]\begin{tabular}{@{}lll}
			\br
			$\epsilon$	& $Q(L=6)$ & $Q(L=12)$ \\ 
			\mr
			$0.350\pi$ & 0.0737 & 0.0047 \\ 
			$0.275\pi$	& 1.8918  & 1.9711\\ 
			\br
		\end{tabular}
	\end{indented}
\end{table} 

\begin{table}[h]
	\caption{\label{table2}Pumped charge $Q$ with $L=6$ for different pump times.}
	\begin{indented}
		\item[]\begin{tabular}{@{}lllll}
			\br
			$\epsilon$ & $Q(T=2000)$ & $Q(T=5000)$& $Q(T=10000)$& $Q(T=20000)$  \\  
			\mr
			$0.350\pi$ &  0.5169 & 0.0968 & 0.0737 & 0.0728  \\ 
			$0.275\pi$	& 1.7001 & 1.8326 & 1.8918 & 1.8988\\ 
			\br
		\end{tabular} 
	\end{indented}
\end{table} 

To illustrate the finite size effects and the adiabatic conditions, we have calculated the pumped charge of the fully occupied bound pair band with $w=1/2$ and $\varphi=\pi/8$ for different system sizes and pump times. We use two paths close to the critical point, namely $\epsilon=0.35\pi$ (trivial) and $\epsilon=0.275\pi$ (non-trivial). In table \ref{table1} the pumped charge $Q$ is calculated for $L=6$, $N=(2,2)$ and $L=12$, $N=(4,4)$. Here $T=10000$ was used to assure that the system was sufficiently adiabatic. This table shows that for $L=12$ the result is closer to the quantized values of 0 and 2 then in the $L=6$ system due to finite size effects. In table \ref{table2} the charge pumped for different values of $T$ is shown for $L=6$, $N=(2,2)$. This data shows that for larger $T$ the result is indeed closer to the quantized values as is expected because the adiabatic conditions are better satisfied. For $L\rightarrow\infty$ and $T\rightarrow\infty$ these values of $Q$ should be precisely quantized to 0 and 2.\\

\section*{References}
\bibliographystyle{unsrt}
\bibliography{library_paper}

\end{document}